\documentstyle[epsfig,12pt,titlepage]{article}
\input epsf

\renewcommand{\theequation}{\thesection.\arabic{equation}}

\outer\def\beginsection#1\par{\medbreak\bigskip
      \message{#1}\leftline{\bf#1}\nobreak\medskip
\vskip-\parskip
      \noindent}

\def\laq{\raise 0.4ex\hbox{$<$}\kern -0.8em\lower 0.62
ex\hbox{$\sim$}}
\def\gaq{\raise 0.4ex\hbox{$>$}\kern -0.7em\lower 0.62
ex\hbox{$\sim$}}
\def\be{\begin{equation}}
\def\ee{\end{equation}}
\def\bees{\begin{eqnarray}}
\def\ees{\end{eqnarray}}

\def \ra {\rightarrow}

\begin{document}
\titlepage
\begin{flushright}
IFUP-TH/53-97 \\
November 1997
\end{flushright}
\vspace{7mm}
\begin{center}
{\bf Anisotropic String Cosmology at Large Curvatures}

\vspace{9mm}

Stefano Foffa and Michele Maggiore\\
{\sl  Dipartimento di Fisica and INFN, piazza Torricelli 2, I-56100 Pisa,
Italy.}

\end{center}
\vspace{9mm}

\vspace{2mm}
\noindent
We study the effect of the antisymmetric tensor field $B_{\mu\nu}$
on the large curvature phase of string cosmology. It is well-known
that a non-vanishing  value of $H=dB$ leads to an anisotropic
expansion of the spatial dimensions. Correspondingly, in the string
phase of the model, including $\alpha '$ corrections, we find
anisotropic fixed points of the evolution, which act as  regularizing
attractors of the lowest order solutions. The attraction basin can also
include isotropic initial conditions for the scale factors. We present
explicit examples at order $\alpha '$ for different values of the
number of spatial dimensions and for different ans\"{a}tze for $H$.

\vfill

\newpage

\renewcommand{\theequation}{1.\arabic{equation}}
\setcounter{equation}{0}
\section {Introduction}
 Cosmological models derived from the low-energy action of string
theory [1-5] can address fundamental
problems of cosmology, like the initial singularity, and at the same
time are a testing ground for our present knowledge of string theory.
The pre-big-bang model developed in~[1-4] has attractive
physical features. It has a scale factor duality
symmetry~\cite{Ven,duality,Meis,KM}  which suggests  the existence of a dual,
``pre-big-bang'' phase and a possible solution of the initial
singularity problem, and has a built-in mechanism for obtaining an
inflationary phase driven by the kinetic energy of the dilaton (see
also~\cite{Lev}). Whether this mechanism is sufficient to solve the
flatness/horizon problems has been a matter of some discussion
recently~\cite{initial}.

Another aspect of the model, which distinguishes it from previous
attempts, is that it allows us to perform quite explicit
computations. This is due to the fact that the evolution starts
in a regime of low curvatures and weak couplings, where
both sigma-model  corrections (controlled by the string constant
$\alpha '=\lambda^2_{s}$) and string loops (controlled by the
value of the dilaton)
are fully under control. The existence of this perturbative regime
allows to perform detailed computation of the spectra of particles
produced  by amplification of vacuum
fluctuations~\cite{spectra}. Important features of these spectra, such
as the $\sim\omega^3$ dependence of the graviton spectrum at low
frequencies, are only sensitive to this phase and therefore are under
good theoretical control.

Starting at the initial time in this perturbative regime, the model
evolves toward a large curvature regime. In this phase $\alpha '$
corrections certainly become important.
In fact, at lowest order $\alpha '$ only enters
as an overall constant in the effective action. Therefore it drops
from the equations of motion and there is no scale at which we can stop
the growth of the curvature. The inclusion of $\alpha '$ corrections
provides such a scale, and changes qualitatively the form of the
solution in the large curvature regime. Computations in this
phase are not under such a good theoretical control. What has been
done to date is the following:

(i) one can study perturbative $\alpha '$
corrections, working at first order in $\alpha '$, i.e. including
terms $\sim R_{\mu\nu\rho\sigma}^2$ in the low energy
action~\cite{GMV}.  Working
at this order we can already see the regularizing effect of $\alpha '$
corrections and we can therefore hope to have a glimpse of
the structure which could be obtained at all orders in $\alpha '$. In
fact, it was found in~\cite{GMV} that a De~Sitter solution with
linearly growing dilaton exists at
all orders in $\alpha '$ if a set of algebraic (rather then
differential) equations admit a real solution.
Such indications, of course, cannot be obtained at zero order in
$\alpha '$, since in this case, as we said, $\alpha '$ drops from the
equations and there is  no scale at which a new regime can set in.

Working at any finite order in $\alpha '$, one must however keep in
mind the limitations of the computation, which are due not only to the
fact that, in the large curvature regime,
higher order corrections are potentially of the same order
as the lowest order terms, but also to the fact that results obtained
at finite order are very sensitive to the scheme used for
renormalization of the sigma model, or equivalently are sensitive to
field redefinitions~\cite{GMV,MM}.

(ii): non-perturbative effects in $\alpha '$ are also very important.
In particular the production of massive string modes is
non-perturbative in $\alpha '$ and provides a mechanism which stops
the growth of the curvature~\cite{MM}.

So, even if the large curvature phase is certainly not as much under
theoretical control as the previous dilaton-dominated phase, still
we can perform a number of computations which gives us some
indications on the behavior of the cosmological model.

Finally, the large curvature phase of the model should be matched with
the standard radiation dominated era. This is the so called graceful
exit problem~\cite{gra}.

In this paper we consider the effect of the antisymmetric tensor field
$B_{\mu\nu}$ on the large curvature phase of the model.
It has been recognized for some time that a non vanishing value  of its
field strength $H=dB$ leads to an anisotropic expansion; the fact that
$H$ is a three-form opens in
principle the possibility that in the presence of a non zero $H$
three spatial dimensions expand while six (or, for bosonic string,
twenty-two )
contract, realizing a
scenario of dynamical ``ten into four'' compactification~\cite{FR}.
These ideas can be tested in the context of the
 model of refs.~[1-4]. The evolution with
a non-vanishing $H$, in the low curvature
dilaton-dominated phase of the model, has   been studied in
ref.~\cite{CLW}, where it has been shown that a three-form $H$
with a component $H_{012}\neq 0$
inevitably leads to an anisotropic cosmology.
The effect of form fields in string cosmology and in M-theory has also
been considered in~\cite{LOW,Lu,PS}.
Here we extend the analysis
to the large curvature region of the model, considering the effect of
perturbative $\alpha '$ corrections. Following the approach
of ref.~\cite{GMV}, we will find the fixed points
of the evolution which
can be obtained after inclusion of perturbative $\alpha '$
corrections,  and we will discuss the resulting scenario for dynamical
compactification of the extra dimensions.

The paper is organized as
follows. In sect.~2 we discuss the model, we recall some of the
relevant results in the literature and
we present possible ans\"{a}tze for the three-form
$H$. In sect.~3 we discuss the corresponding solutions of the
equations of motion.

\renewcommand{\theequation}{2.\arabic{equation}}
\setcounter{equation}{0}

\section { The model}

The starting point of our investigation is the effective action, up to
${\cal O}(\alpha ')$, of the bosonic sector of a string theory 
with nontrivial
antisymmetric tensor field strenght $H_{\rho\mu\nu}=\partial_{[\rho}
B_{\mu\nu]}=\partial_{\rho}B_{\mu\nu}+
\partial_{\mu}B_{\nu\rho}+\partial_{\nu}B_{\rho\mu}$;
in the string frame it has the following form~\cite{Mat}:
\bees
S_{0} 
&=& - \frac{1}{2 \lambda^{d-1}_{s}} \int d^{d+1} x \sqrt{|g|}e^{-\phi}
\left\{ R+\left(\nabla\phi\right)^2-
\frac{H^2}{12}-\frac{k\alpha '}{4}\left[R^{\mu\nu\rho\sigma}
R_{\mu\nu\rho\sigma}+ \right. \right. \nonumber\\
\label{S0}
& & \left. \left. - \frac{1}{2}R^{\mu\nu\rho\sigma}H_{\mu\nu\alpha}
{H_{\rho\sigma}}^{\alpha}+
\frac{1}{24}H_{\mu\nu\lambda}{H^{\nu}}_{\rho\alpha}
H^{\rho\sigma\lambda}{H_{\sigma}}^{\mu\alpha}
- \frac{1}{8}\left(H^2_{\mu\nu}\right)^2\right]\right\}\ ,\label{1}
\ees
where $k=1,\frac{1}{2},0$ for bosonic, heterotic and type II strings
respectively. We restrict in the following to the bosonic and
heterotic string case. For the type II string the first corrections
are due to terms $\sim R_{\mu\nu\rho\sigma}^4$~\cite{GW}.

Our conventions are $\eta_{\mu\nu}=(+, - , -, -, \dots)$ and
${R^{\mu}}_{\nu\rho\sigma}=
(\partial_{\rho}\Gamma^{\mu}_{\nu\sigma}-\dots)$. In eq.~(\ref{S0}) we
have defined
$H_{\mu\nu}^2=H_{\mu\alpha\beta}{H_{\nu}}^{\alpha\beta}$ and
$H^2=H_{\mu\nu\rho}H^{\mu\nu\rho}$.
The $\cal O(\alpha ')$ part of (\ref{1}) is not uniquely fixed: 
actually, by
means of the  redefinitions of the fields $g_{\mu\nu}\ra g_{\mu\nu}
+\delta g_{\mu\nu}, \phi\ra\phi +\delta\phi, B_{\mu\nu}\ra
B_{\mu\nu}+\delta B_{\mu\nu}$ with
\bees
\delta g_{\mu\nu} &=& k\alpha' \left\{b_{1} R_{\mu\nu} + b_{2} 
\partial_{\mu}
\phi\partial_{\nu}\phi + g_{\mu\nu} \left[b_{2} \left(\nabla\phi\right)^2 +
b_{4} R + b_{5} \Box\phi + b_{6} H^2\right] + b_{7} H^2_{\mu\nu}\right\}\ ,
\nonumber\\
\delta \phi &=& k\alpha'\left[c_{1}R+c_{2}\left(\nabla\phi\right)^2
+ c_{3}\Box\phi + c_{4}H^2\right]\ , \nonumber\\
\delta B_{\mu\nu} &=& k\alpha'\left[ \left(d_{1}\nabla^{\rho}+ d_{2}\partial
^{\rho}\phi\right)H_{\rho\mu\nu}\right]\ , \nonumber
\ees
one can generate  an action of the type
\bees
S &=& S_{0} + \frac{1}{2 \lambda^{d-1}_{s}}
\left(\frac{k\alpha '}{4}\right)
\int d^{d+1} x \sqrt{-g}e^{-\phi}
\left[a_{1}R_{\mu\nu}R^{\mu\nu} + 
a_{2} R^2 + a_{3} \left(\nabla\phi\right)^4
+ a_{4}R^{\mu\nu}\partial_{\mu}\phi\partial_{\nu}\phi + \right. \nonumber\\
& & + a_{5} R\left(\nabla\phi\right)^2 +  a_{6} R\Box\phi + a_{7}\Box\phi
\left(\nabla\phi\right)^2 + a_{8}\left(\Box\phi\right)^2 + a_{9}R H^2 +
a_{10}R^{\mu\nu}H^2_{\mu\nu} + \nonumber\\
& & a_{11}H^2_{\mu\nu}\partial^{\mu}\phi\partial^{\nu}\phi + a_{12} \left(
\nabla^{\mu}\partial^{\nu}\phi\right)H^2_{\mu\nu} + a_{13} H^2 \left(\nabla
\phi\right)^2+ a_{14} H^2 \Box\phi+ a_{15} \left(H^2\right)^2 + \nonumber\\
& & \left. a_{16}\left(H^2_{\mu\nu}\right)^2 + a_{17}\left(\nabla_{\rho}
H^{\rho\mu\nu}\right)\left(\nabla^{\sigma}H_{\sigma\mu\nu}\right) +
a_{18}\left(\nabla_{\rho}H^{\rho\mu\nu}\right)\partial^{\sigma}\phi
H_{\sigma\mu\nu}\right]\ , \label{18}
\ees
where $S_0$ is given by eq.~(\ref{S0}) and
we have eliminated terms that can be reduced to those
displayed, by means of integration by parts or use of Bianchi
identity.
The parameters $a_{i}$ are functions of $b_{i},c_{i},d_{i}$
and satisfy the following relations:
\bees
& &a_{2}+a_{3}-a_{5}-a_{6}+a_{7}+a_{8}=0\ , \nonumber\\
& &a_{1}+4a_{10}-4a_{12}=0\ , \nonumber\\
& &a_{12}+4a_{16}=0\ , \\
& &a_{1}-a_{4}-4a_{11}-4a_{17}-a_{18}=0\ , \nonumber\\
& &5a_{1}+9a_{2}+4a_{3}-2a_{4}+6a_{5}+36a_{9}+24a_{13}+144a_{15}=0\ ,
\nonumber\\
&
&5a_{1}+25a_{3}-5a_{4}+15a_{7}+9a_{8}+60a_{13}+36a_{14}+144a_{15}=0\, .
\nonumber
\ees
Within these constraints the $a_{i}$ can be chosen at will with the
appropriate field redefinitions: the different actions one obtains all
reproduce the correct string theory $S$-matrix elements on-shell and
so they are equally all
good candidates for a description of
the low energy regime (up to ${\cal O}
(\alpha ')$) of the corresponding string theory. The problem is that, since
we work at a finite order in $\alpha '$, different schemes ({\it i.e.}:
different choices of the $a_{i}$) can
lead to different physical results as, for instance, the existence or not of
a fixed point of the evolution. We will return to this question later, but
from the above consideration it is clear that whatever result is
obtained within a particular field redefinition should be taken only as an
illustrative example rather than a definite prediction.

After this remark, let us  make our choice for the $a_{i}$,
in order to have a concrete model to deal with. First of all, we require
the action (\ref{18}) not to produce higher than second order
derivatives in the
equations of motion. This can be done by allowing only the presence
of the structures reported in eq.$(4.4)$ of \cite{Meis}; this
request reduces the $18-$parameter family of eq. (\ref{18}) to the 
following
$2-$parameter one:
\bees
S &=& - \frac{1}{2 \lambda^{d-1}_{s}} \int d^{d+1} x \sqrt{|g|}e^{-\phi}
\left\{ R+\left(\nabla\phi\right)^2-
\frac{H^2}{12}-\frac{k\alpha '}{4}\left[R^{\mu\nu\rho\sigma}
R_{\mu\nu\rho\sigma} + \right. \right. \nonumber\\
& & - \left(1+\lambda_{1}-\lambda_{2}\right)\left(\nabla\phi\right)^4+
\lambda_{1}\Box\phi\left(\nabla\phi\right)^2
+ \lambda_{2}\left(g^{\mu\nu}R - 2 R^{\mu\nu}\right)\partial_{\mu}\phi
\partial_{\nu}\phi + \nonumber\\
& & - \frac{1}{2}\left(R^{\mu\nu\rho\sigma}H_{\mu\nu\alpha}
H_{\rho\sigma}^{\alpha} - 2R^{\mu\nu}H^2_{\mu\nu}+\frac{1}{3}RH^2\right) +
\nonumber\\
& & + \frac{1}{24}H_{\mu\nu\lambda}{H^{\nu}}_{\rho\alpha}
H^{\rho\sigma\lambda}{H_{\sigma}}^{\mu\alpha} +
\frac{5}{144}\left(H^2\right)^2 - \frac{1}{8}\left(H^2_{\mu\nu}\right)^2
\nonumber\\
& & \left. \left.- \left(1- \frac{\lambda_{2}}{2}\right)H^2_{\mu\nu}
\partial^{\mu}
\phi\partial^{\nu}\phi+\frac{2}{3}\left(1+\frac{\lambda_{1}}{4}-
\frac{7\lambda_{2}}{8}\right)H^2 \left(\nabla\phi\right)^2\right]
\right\}\ .
\label{model}
\ees
Moreover, we set $\lambda_{1}=\lambda_{2}=0$; with this prescription
our  model is 
the straightforward generalization of the one studied in ref. \cite{GMV}
to the case of nonvanishing antysimmetric tensor field strenght.

Since the basic ingredients of our model are:

(i) the presence of $\alpha '$ corrections,

(ii) a nontrivial antisimmetric tensor field strenght $H_{\mu\nu\rho}$,\\
we first briefly recall what are the principal effects that these
ingredients produce when they act separately.\\
Concerning the role of $\alpha '$ corrections alone, we refer to the
treatment  in ref.~\cite{GMV}, since in the next section we will apply
quite the same procedure to the case of nonvanishing $H_{\mu\nu\rho}$.
The model considered in \cite{GMV} corresponds to the ours without
antisymmetric tensor; by means of the following ansatz for the metric
\be
g_{00}=N^2(t),\ \ \ \ \ \ g_{ij}= -\delta_{ij}e^{2\beta(t)},
\ \ \ \ \ \ i,j=1, \dots , d\ , \nonumber
\ee
and after integration by parts, the actions becomes (discarding
an irrelevant overall multiplicative constant)
\be
S = \int dt e^{d\beta-\phi}\left[\frac{1}{N}\left(\dot{\phi}^2 +
d(d-1)\dot{\beta}^2 - 2d\dot{\beta}\dot{\phi}\right) 
-\frac{k\alpha '}{4N^3}
\left(c_{1}\dot{\beta}^4 + c_{2}\dot{\phi}\dot{\beta}^3 -
\dot{\phi}^4\right)\right], \nonumber
\ee
where
\be
c_{1}=-
\frac{d}{3}(d-1)(d-2)(d-3),\ \ \ c_{2}=\frac{4d}{3}(d-1)(d-2). \nonumber
\ee
Varying the action with respect to $\beta$, $\phi$ and $N$, one gets two
dynamical equations of motions plus a constraint on the initial data.
To look for a solution one makes the ansatz
$\dot{\phi}=x=$ const, $\dot{\beta}=y=$ const; then, in the gauge $N=1$,
the differential equations reduce
to three algebraic equations in two unknowns.
One should not be worried about this fact, because, due to
reparametrization invariance, these equations are not 
independent, {\it i.e.}
once the constraint and one of the ``dynamical'' equation is satisfied,
the other is automatically satisfied.
Therefore the problem of finding a solution in the string phase is reduced
to solving the following system of two nonlinear equations in two unknowns:
\bees
x^2 &+& d(d-1)y^2- 2dxy - \frac{k\alpha '}{4}\left(c_{1}y^4 + c_{2}xy^3
-x^4\right)+ \nonumber\\
&-& (dy-x)\left[-2x + 2dy + \frac{k\alpha '}{4}
\left(c_{2}y^3 - 4x^3\right)\right]=0, \nonumber\\
x^2 &+& d(d-1)y^2 -2dxy -\frac{3k\alpha '}{4}\left(c_{1}y^4 + c_{2}xy^3
-x^4\right)=0. \label{sys}
\ees
In \cite{GMV} it was verified that the system (\ref{sys}) has a real solution
for any $d$ from $1$ to $9$. Moreover, by integrating numerically the full
differential equations for $\beta$ and $\phi$ and imposing the constraint on
the initial data, it was found that this solution acts as late-time
attractor for the evolution of the system, i.e. is a fixed point of
the evolution, and its basin of attraction
includes initial conditions corresponding to a state of pre-big bang evolution
from the vacuum.
The idea that emerges from the above example is that large curvature
corrections to the lowest order string effective action can regularize the
otherwise singular pre-big bang solution.
As previously mentioned,
it must be remembered that the last statement depends
on the choice of the field redefinition:  it has been verified
(\cite{GMV, MM}) that there are other choices of the $a_{i}$ for which the
above fixed point are not smootly connected to the perturbative vacuum or do
not exist at all. Nevertless, it can still be
useful to study models with corrections
${\cal O}(\alpha ')$,  at least to obtain some
indicative examples of what can
be the role of large curvature corrections in pre-big bang cosmology.

If perturbative $\alpha '$ corrections, computed at all orders in
$\alpha '$, would not regularize the lowest order solution, then the
regularizing mechanism is the production of massve string modes, which
is non perturbative in $\alpha '$, and has been discussed in
ref.~\cite{MM}.

Concerning the action of the antisymmetric tensor on the low curvature
part of our model ({\it i.e.} without $\alpha '$ corrections), it
has been studied
in detail by Copeland et al. (see ref. \cite{CLW} and references therein):
they present the explicit
solutions of the lowest-order equations of motion in $D=4$ and find that a
nonvanishing $H_{\mu\nu\rho}$ noticeably affects the dynamics of the system.
In particular they find
that a homogeneous $B_{ij}$, which corresponds to
a nonvanishing $H_{0ij}$ (latin indices
indicate spatial components), produces anisotropic evolution even in the
presence of isotropic initial conditions. They also studied the case
$H_{0\mu\nu}=0,H_{ijk}\neq0$ in $4+n$ dimensions: it turns out that the
presence of the antisymmetric tensor accelerates the expansion of
three spatial directions with respect to the other $n$, leading
to an  anisotropic $3+n$ cosmology. However, without $\alpha '$
corrections, the evolution eventually runs into a singularity.

The fact that nonvanishing form fields produce an anisotropic expansion
is quite general: the presence of the field strenght $F_{r}$ of
an $r-$form in a string effective action makes the system evolve in an
anisotropic way and (see also~\cite{FR}) the
number of spatial dimensions that are separated from the other is equal to the
number of spatial components of the field strenght; so, given an $r-$form, two
simple situations are possible:

a)$F_{i_{1}\dots i_{r+1}}\neq0$,$F_{0i_{1}\dots i_{r}}=0$. In this situation
$r+1$ spatial dimensions separate from the other;

b)$F_{0i_{1}\dots i_{r}}\neq0$,$F_{i_{1}\dots i_{r+1}}=0$. In this case the
splitting is $r$ vs. $D-r-1$.\\
We shall refer to a) and b) as the {\it solitonic}
and {\it elementary} ans\"{a}tze, respectively: the nomenclature
comes from the relation with elementary and solitonic $p$-branes in
M-theory explored in ref.~\cite{LOW}.

Different types of string effective action have a different content of
form fields. However, the three form $H=dB$ is common to all of them
(bosonic, heterotic, type II) and we focus on it in the following.

\section{The evolution with a non-vanishing $H$}
Having recalled what are the basic features of models caracterized by
torsion and large curvature corrections separately, we want to study what
happens when both  are present. We will show that in this case
the distinctive features of each factor are not lost, but rather merge in the
way one expects: the presence of anisotropic fixed points.\\
We are going to apply the procedure of ref. \cite{GMV} to our model in order
to find the fixed points of the evolution; obviously we must release the
request of isotropy, but we still want to work with a homogeneous metric,
so we assume that all fields depend only on time.\\
Let us work first in $D=4$ and consider the elementary ansatz: with a
spatial rotation we can always reduce ourselves to the situation in which
only one  component $H_{0ij}$ of the field strenght,
 say $H_{012}$ (and its permutations),
 is nonzero and we call it simply $h_{e}$. The equations that
$h_{e}$ must obey are the integrability condition $dH=0$ and the
${(1,2)}$--component  of the equation of motion for $B_{\mu\nu}$
(the other components are trivially satisfied by our ansatz and by the
request of homogeneity):
\bees
\frac{\delta S}{\delta B_{\mu\nu}}=- \nabla^{\rho}\frac{\delta S}
{\delta \partial_{\rho}B_{\mu\nu}}=0.\ \label{Bmunu}
\ees
The integrability condition reduces to
\bees
\partial_{3}h_{e}=0, \nonumber
\ees
which is obviously satisfied by an homogeneous field.
With our choice for $h_{e}$ we
expect directions $1$, $2$ to evolve differently from
direction $3$, and so we describe
the metric tensor in the following way:
\bees
g_{00}=N^2(t),\ \ \ \ g_{11}=g_{22}=-e^{2\beta(t)},\ \ \ \
g_{33}=-e^{2\gamma(t)}, \ \ \ \ g_{\mu\nu}=0\ {\rm if}\ \mu\neq\nu . 
\nonumber
\ees
We can now  compute all the terms contained in action
(\ref{model}), and we get
\bees
S &=& \int dt\, e^{2\beta+\gamma-\phi}\left\{\frac{1}{N}
\left[\dot{\phi}^2 +
2\dot{\beta}\left(\dot{\beta} + 2\dot{\gamma}\right) - 2\dot{\phi}\left(
2\dot{\beta}+\dot{\gamma}\right)-\frac{1}{2}e^{-4\beta}h_{e}^2\right]
\right. +
\nonumber\\
& & \left. -\frac{k\alpha '}{4N^3}
\left[8\dot{\phi}\dot{\beta}^2\dot{\gamma} -\dot{\phi}^4 +
2e^{-4\beta}\dot{\phi}^2 h_{e}^2\right]\right\}. \nonumber
\ees
It is now straightforward to write down the equations of motion; we report
them  introducing a new variable $z\equiv e^{-2\beta}h_{e}$:
\bees
&&- \dot{\phi}^2 - 2\dot{\beta}\left(\dot{\beta}+2\dot{\gamma}\right)
+ 2\dot{\phi}\left(2\dot{\beta} + \dot{\gamma}\right) +
\frac{3k\alpha '}{4} \left(8\dot{\phi}\dot{\beta}^2\dot{\gamma} -
\dot{\phi}^4\right)+ \frac{z^2}{2}\left(1+ 3k\alpha '
\dot{\phi}^2\right)=0, \nonumber\\
&&- 2\ddot{\phi} + 2\left(2\ddot{\beta} + \ddot{\gamma}\right) +
\dot{\phi}^2 + 2 \left(3\dot{\beta}^2 + \dot{\gamma}^2 +
2\dot{\beta}\dot{\gamma}\right) - 2\dot{\phi}\left(2\dot{\beta} +
\dot{\gamma}\right) + \nonumber\\
&&+ \frac{k\alpha '}{4} \left[3\dot{\phi}^4 - 4\dot{\phi}^3
\left(2\dot{\beta} + \dot{\gamma}\right) + 8\dot{\beta}\left(2\ddot{\beta}
\dot{\gamma} + \ddot{\gamma}\dot{\beta}\right)
- 12\ddot{\phi}\dot{\phi}^2 + 8\dot{\beta}^2 \dot{\gamma}
\left(2\dot{\beta} + \dot{\gamma}\right)\right]
+ \nonumber\\
&&+ \frac{z^2}{2}\left[1 + k\alpha ' \left(-\dot{\phi}^2 +
2\dot{\phi}\left(2\dot{\beta} + \dot{\gamma}\right)
+2 \ddot{\phi}\right)\right]+k\alpha ' z\dot{z}\dot{\phi}=0, \nonumber\\
&&2\ddot{\phi} - 2\left(\ddot{\beta}+\ddot{\gamma}\right) - \dot{\phi}^2
 - 2\left(\dot{\beta}^2 + \dot{\beta}\dot{\gamma} +
\dot{\gamma}^2\right) +
2\dot{\phi}\left(\dot{\beta} + \dot{\gamma}\right) + \nonumber\\
&&+ \frac{k\alpha '}{4}\left[\dot{\phi}^4
+ 8\dot{\phi}\dot{\beta} \dot{\gamma}\left(\dot{\beta} +
\dot{\gamma}\right) + 8\dot{\phi}\left(\ddot{\beta}\dot{\gamma} +
\ddot{\gamma}\dot{\beta}\right) + 8\left(\ddot{\phi}-
\dot{\phi}^2\right)\dot{\beta}\dot{\gamma}\right] +
\frac{z^2}{2}\left(1 + k\alpha ' \dot{\phi}^2\right)=0, \nonumber\\
&&2\ddot{\phi} - 4\ddot{\beta} - \dot{\phi}^2 - 6\dot{\beta}^2 +
4\dot{\phi}\dot{\beta} + \frac{k\alpha '}{4}\left[\dot{\phi}^4 +
16\dot{\phi}\dot{\beta}^3 + 16\dot{\phi}\dot{\beta}\ddot{\beta} +
8\left(\ddot{\phi}-\dot{\phi}^2\right)\dot{\beta}^2\right] + \nonumber\\
&&- \frac{z^2}{2}\left(1 + k\alpha ' \dot{\phi}^2\right)=0, \label{eq1}
\ees
while eq.(\ref{Bmunu}) becomes
\bees
\dot{z}\left(1 + k\alpha ' \dot{\phi}^2\right) + z\left[\left(\dot{\beta}-
\dot{\phi}\right)\left(1 + k\alpha ' \dot{\phi}^2\right)+
2k\alpha ' \dot{\phi}\ddot{\phi}\right]=0. \label{B1}
\ees
By setting $\dot{\phi}=x=$ const, $\dot{\beta}=y=$ const,
$\dot{\gamma}=w=$ const, $z=$ const (note that this does not mean
$h_{e}=$ constant), we obtain from (\ref{eq1}), (\ref{B1})
an algebraic system of five equations in four unknowns; as previously 
stated, the equations 
are not independent and we have numerically solved the system,
finding the following fixed points (we have set $k\alpha '=1$):\\
\begin{center}
\begin{tabular}{|c|c|c|c|c|}
\hline
 & $x$ & $y$ & $w$ & $z$ \\
\hline
\hline
$A$ & $1.40439$ & $0.616596$ & $0.616596$ & $0$ \\
\hline
$B$ & $1.41421$ & $0.707107$ & $0.471405$ & $0$ \\
\hline
$C$ & $1.66003$ & $0.378784$ & $1.66003$ & $0.442278$ \\
\hline
\end{tabular}
\end{center}
We see that $A$ is the same (isotropic) fixed point found in \cite{GMV}.
Its existence follows from the fact that for $z=0$ the above equations
reduce to the equations of ref.~\cite{GMV}. Studying its domain of
attraction
by direct numerical integration of the equations of motion
we found that $A$ can be a late-time attractor only if we impose $z=0$
as initial condition (this condition is conserved by eq. (\ref{B1})).

Quite the same statement holds for the anisotropic fixed point $B$, i.e.
it doesn't attract solutions of the equations of motion whose initial
conditions are such that $z\neq0$, while it is the late time attractor
of a certain region of the $(x,y,w,z)-$space characterized,
as well as by $z=0$, also by $y\neq w$: we conclude that in both cases
the antisymmetric tensor plays little role and that the anisotropy
of $B$ is uniquely due to the choice of anisotropic initial conditions.\\ 
Up to now we are only elaborating on the results of
ref.~\cite{GMV}, in the sense
that we have found that even non-isotropic fixed points are possible,
and that the presence of torsion seems to compromise the stability
of the system, i.e., in the enlarged parameter space which includes $z$
these fixed point have an attraction basin of  zero  measure.
On the contrary, $C$ is the first real novelty of our analysis:
since $z_{C}\neq0$, it is clear that its existence is due to the presence
of the antisymmetric tensor; moreover, it turns out to be a late-time
attractor of a region of the phase space with non-zero measure.
Unfortunately, we have found by numerical integration that
this region does not contain isotropic ($y=w$) initial conditions.
One can hope that working in a different number of dimensions could
improve the situation: we will see that this can indeed be the case,
but we prefer to develop this idea by studying the solitonic ansatz that,
on account of what stated at the end of the last section, should provide a
more appealing $3+n$ dimensional separation of spatial dimensions,
rather than the $2+(n+1)$
separation provided by the elementary ansatz. \\
We then choose  $H_{123}\equiv h_{s}$ as the only non-zero
component of $H_{\mu\nu\rho}$, modulo permutations of
the indices; the integrability condition now reads
\bees
\partial_{\mu}h_{s}=0, \nonumber
\ees
where the index $\mu$ runs from $0$ to $n+3$ and $\mu\neq 1,2,3$.
The equations with $\mu =4,\ldots n+3$
are trivially satisfied on account of
homogeneity, while the equation with $\mu =0$
implies that $h_{s}$ must be a
constant; moreover, our ansatz automatically satisfies every component
of the equation of motion (\ref{Bmunu}).\\ 
After having parametrized the metric tensor in the following way
\bees
g_{00}=N^2,\ \ \ \ g_{ij}=-\delta_{ij}e^{2\beta},\ \ \ i,j=1,\dots ,3\ ,
\ \ \ \ g_{ab}=-\delta_{ab}e^{2\gamma},\ \ \ a,b=4,\dots ,n+3\ , \nonumber
\ees
it is not difficult to write down the new form of action (\ref{model}):
\bees
&S& = \int dt e^{3\beta+n\gamma-\phi}\left\{\frac{1}{N}\left[\dot{\phi}^2 +
6\dot{\beta}^2
+6n\dot{\beta}\dot{\gamma}+n(n-1)\dot{\gamma}^2
- 2\dot{\phi}\left(3\dot{\beta}+n\dot{\gamma}\right)+
\frac{N^2}{2}e^{-6\beta}h_{s}^2\right]\right. + \nonumber\\
&-&\frac{k\alpha '}{4N^3}
\left[-8n\dot{\beta}^3\dot{\gamma}-12n(n-1)\dot{\beta}^2\dot{\gamma}^2-
4n(n-1)(n-2)\dot{\beta}\dot{\gamma}^3-\frac{n}{3}(n-1)(n-2)(n-3)
\dot{\gamma}^4+ \right. \nonumber\\
&&\ \ \ \ \ \ \ \ +\dot{\phi}^3\left(8\dot{\beta}^3+24n\dot{\gamma}
\dot{\beta}^2+12n(n-1)\dot{\gamma}^2\dot{\beta}+\frac{4n}{3}(n-1)(n-2)
\dot{\gamma}^3-\dot{\phi}^3\right) + \nonumber\\
&&\ \ \ \ \ \ \ \  +\left. \left. N^2e^{-6\beta}h_{s}^2\left(n(n-1)
\dot{\gamma}^2-6n\dot{\gamma}\dot{\beta}-2n\dot{\gamma}\dot{\phi}-
4\dot{\phi}^2\right)\right]\right\}. \nonumber
\ees
We are now ready to repeat the usual procedure of writing down the equations
of motion (this time we will not report them), finding the fixed points,
and integrating the full numerical system, for generic values
of $n$.\\
The case $n=0$, {\it i.e.} four dimensional space-time,
doesn't tell anything interesting: we know that, since we have
only three spatial dimension, $h_{s}$ cannot induce any anisotropy on
the sistem: actually the only fixed point we found is the
fixed point $A$ of the
previous discussion, with the difference that this time $A$ is a good
attractor even if we start with $z\neq0$ (having defined a new
$z\equiv e^{-3\beta}h_{s}$).\\
Trying different values for $n$ we have found, as well as an
 isotropic fixed point for each $n$, also many anisotropic ones.
Unfortunately, they result to be all (with reference to the previous
discussion) $B$-like, {\it i.e.} not torsion-induced fixed point, as well as
unstable; but there is one exception, for $n=2$: a fixed point $F$
of coordinates
\begin{center}
 $x=1.38259,\ y=0.238441,\ w=0.438904,\ z=0$.
\end{center}
In fact the stability analysis reveals not only that $F$ is a good attractor,
but also that the attraction basin includes even an isotropic region,
characterized by $h_{s}\neq0$; in other words, the nonvanishing
component of the antisymmetric field strenght drives an initially isotropic
region towards an anisotropic fixed point. The behaviour of the system
is presented in fig.~$1$.
\begin{figure}
\centerline{\psfig{figure=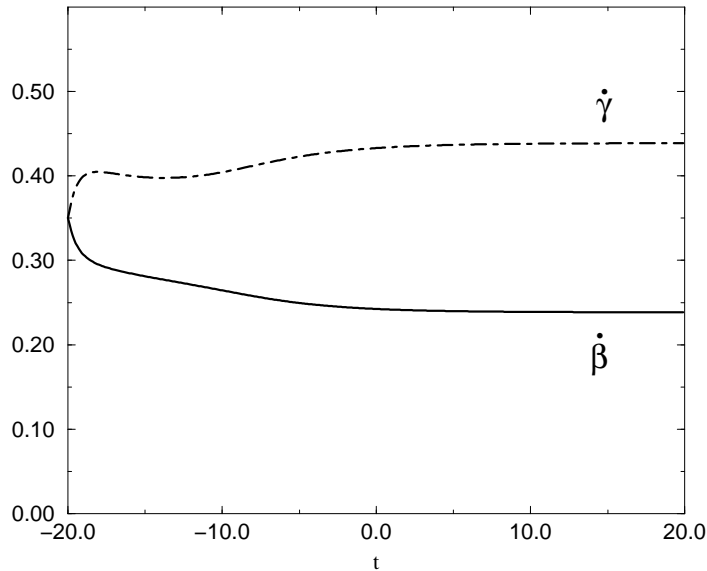,height=9cm}}
\caption{Evolution of the system towards anisotropic fixed point 
$F$, starting
from isotropic initial conditions.}
\end{figure}
As we already discussed, our model has only an illustrative value.
Therefore, we are not too much disappointed
by the fact that the compactification showed in $F$
acts in the ``wrong'' sense (three dimensions expanding {\it slower} than
the others): we think in fact that this behaviour is not a general
 rule. Our
feeling is enforced by some trials we have made with various values of $n$:
although 
we have not found significant fixed points, we have seen that in many
cases, starting from isotropic initial conditions, three dimensions expand
driven by the torsion while the others contract (see fig.~$2$  for
an example in $D=10$).
Parenthetically, we note that ``contraction'' and ``expansion'',
{\it i.e.} the signs of $\dot{\beta}$ and $\dot{\gamma}$,
are frame-dependent concepts: 
in fact, since the metric tensor and time in the Einstein frame
(here denoted by a tilde) are related to those of the string frame
(which is by definition the frame in which the action has the
form~(\ref{S0}))   by
\bees
{\tilde{g}}_{\mu\nu}={\rm exp}\left(-\frac{2\phi}{D-2}\right)g_{\mu\nu},
\ \ \ \ 
\frac{d\tilde{t}}{dt}
={\rm exp}\left(-\frac{\phi}{D-2}\right), \label{frames1}
\ees
then the Hubble parameter in the $E$-frame can be written as
\bees
\dot{\tilde{\beta}}=e^{\frac{\phi}{D-2}}\left(\dot{\beta} -
\frac{1}{D-2}\dot{\phi}\right). \label{frames2}
\ees
Equations (\ref{frames1}), (\ref{frames2}) show that what is really 
frame-independent is
the ratio of the scale factors;
it follows that the compactification process, that is the growth of some
dimensions {\it with respect} to others, is not altered
by the choice of the frame.
As discussed in detail in ref.~\cite{GV2}, statements concerning the
scale factors are in general frame-dependent. A scale factor growing
in the string frame can be decreasing in the Einstein frame, but
physical properties like the number of e-folds of an inflationary
period, or the spectrum of metric perturbations amplified in the
course of the background evolution, are independent of the frame.
\\
For the same reason previously discussed, we are not especially
 worried about the fact
that we have not found appealing attractors in a more significant number of
dimensions, say $10$ or $26$, (even if we would had liked it!); it is
possible  that
the system exibits the expected behaviour for another choice of the $a_{i}$
(we have unsuccessfully tried some of it, but we explored only a small
fraction of the whole space of parameters). In other words, it is
clear that  a model in which $\alpha '$ corrections are
truncated to any finite order cannot be used to
 {\it prove} anything; what we rather could hope was to find
(as we did it) some concrete example of the action of ${\cal O}(\alpha ')$
corrections and antisymmetric tensor combined togheter, {\it i.e.}
non singular,  anisotropic cosmological solutions.\\
\begin{figure}
\centerline{\psfig{figure=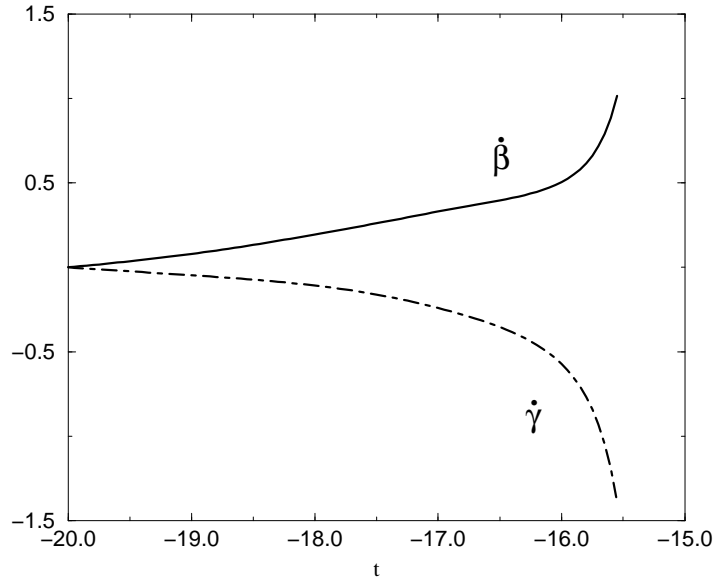,height=9cm}}
\caption{Behaviour of the system in $D=10$, starting from isotropic initial
conditions.}
\end{figure}
Finally, another possibility for obtaining regular anisotropic
solutions, is that in the low curvature regime  three
spatial dimension expand and the remaining contract, as can be
obtained easily with the solitonic ansatz,  and then
 the regularization mechanism is  provided by massive string
modes production, as discussed in ref.~\cite{MM} (see
ref.~\cite{Par} for particle production in anisotropic space-times).
In this case we do not expect a fixed point describing a DeSitter
phase but rather a bounce in the scale factor, as it happens in the
isotropic case.

\end{document}